\documentclass[reqno,11pt]{amsart}
\usepackage{color,xcolor}
\usepackage{amssymb}
\newtheorem{theorem}{Theorem}[section]
\newtheorem{hyp}{Hypothesis} 
\newtheorem{proposition}[theorem]{Proposition} 
\newtheorem{corollary}[theorem]{Corollary} 
\usepackage{enumerate}
\usepackage{enumitem}
\usepackage{tikz}
\usetikzlibrary{math} %needed tikz library
\usetikzlibrary{calc}
\usepackage{pgfplots}
\usetikzlibrary{arrows.meta}
\pgfplotsset{compat=1.13}
\usepgfplotslibrary{fillbetween}
\usetikzlibrary{intersections, pgfplots.fillbetween}
\pgfdeclarelayer{bl}
\pgfdeclarelayer{or}
\pgfsetlayers{bl,main,or}
\usepackage{tabularx}
\usepackage[font={small,stretch=0.80}, labelfont=bf,textfont=it]{caption}
\usepackage{mathtools}%
\mathtoolsset{showonlyrefs}%
\usepackage{hyperref}
\begin{document}

\title[Number of points and symmetries of starbursts]{An explanation of the number of points and symmetries of starbursts}

\author[S. Barbero, A. M.~Delgado and L.~Fern\'{a}ndez]{Sergio Barbero, Antonia M.~Delgado, and Lidia Fern\'{a}ndez}

\address[S. Barbero]{Instituto de \'Optica (CSIC), Madrid (Spain)}
\email{sergio.barbero@csic.es}

\address[A. M. Delgado]{Instituto de Matem\'aticas IMAG \&
Departamento de Ma\-te\-m\'{a}\-ti\-ca Aplicada, Facultad de Ciencias. Universidad de Granada (Spain)}
\email{amdelgado@ugr.es}

\address[L. Fern\'andez]{Instituto de Matem\'aticas IMAG \&
Departamento de Ma\-te\-m\'{a}\-ti\-ca Aplicada, Facultad de Ciencias. Universidad de Granada (Spain)}
\email{lidiafr@ugr.es}

\begin{abstract}
Starbursts are the light intensity patterns seen when small bright sources are looked at night, typically stars. Starburst shapes are produced when the presence of the eye's wave aberrations generates caustics (light concentration) at the retina. A fascinating, but never explained fact about starbursts is that they usually present a $p$-fold symmetry pattern. We provide a theoretical explanation of the number of points and symmetries of starbursts, based on the geometric and algebraic properties of the wave aberration function expressed as a Zernike polynomial expansion. Specifically, we investigate the number and distribution of saddle cusps of Gauss of the Hessian of the wave aberration function. We also establish the connections between those points with the symmetries and the number of starburst points. We found that starbursts are likely generated by axially symmetric dominated wave aberrations with some amount of non-axially symmetric terms. For instance, whereas a wave aberration with a dominant spherical aberration (Zernike polynomial $Z_4^{0}$) plus $Z_3^{3}$ may induce a $3$ points starburst with a $3$-fold symmetry, a wave aberration combining $Z_4^{0}$ and $Z_4^{4}$ may induce a $4$-fold symmetry starburst with $4$ or $8$ points.
\end{abstract}

\maketitle

\smallskip


%%%%%%%%%%%%%%%%%%%%%%%%%%  body  %%%%%%%%%%%%%%%%%%%%%%%%%%
\section{Introduction}
When looking at a star, under low-light conditions and skies free from excessive artificial light emissions~\cite{Bara:21}, most people perceive some structured bright patterns, which have been called \textit{starbursts}. Nowadays, it is recognized that the origin of starbursts relates to caustic formation at the retina surface~\cite{Rubinstein19, Xu}.

Optically, a star in the firmament can be modeled (geometrical optics framework) as a light punctual source from which a line congruence, \emph{i.e.}, a family of rays with an associated orthogonal surface (the \textit{wavefront}), emerges. After refraction within the eye, this wavefront focuses on the retina, generating a caustic pattern instead of a perfect focal point because the eye is far from ideal, introducing what is called \textit{optical aberrations}. 
Caustics can be mathematically defined --there are different definitions-- as the loci of the centers of curvature of the wavefront~\cite{Stavroudis:1972}; in other words, caustics are the spatial locations where two or more rays intersect, hence providing those locations with a higher concentration of light. So far, we have mentioned only geometrical optics, but the full detailed explanation of such caustic shapes ultimately involves wave optics~\cite{Rubinstein19}. Nevertheless, if one concentrates on the symmetry properties of the caustic patterns, geometrical optics is enough. 

The starbursts are only perceived when the eye is affected by a noticeable amount of aberrations~\cite{Rubinstein19, Xu}. These are mathematically described as the \textit{wave aberration function}, defined as the difference between the actual wavefront and the ideal reference spherical wavefront that would eventually concentrate all rays in a single point.  Considering that the angular size of a typical star is much smaller than the maximum resolution of the eye, namely, $1$ arcmin, the image of a star at the retina can only be resolved if the amount of wave aberration generates caustics that exceed that threshold resolution value of $1$ arcmin~\cite{Navarro:97}.

Starburst patterns can be very diverse~\cite{Xu}; still, some very typical ones are those in which a bright central area is surrounded by clearly marked intensity spikes, from now on called: \textit{starburst points}. We call these starbursts \textit{star-like}, which is not a redundant terminology as there are starbursts that are perceived differently, for instance, as a snowflake pattern~\cite{Xu}. 

A striking fact of perceived starbursts is that they usually present a more or less clear $p$-fold symmetry pattern, \emph{i.e.}, it can be mapped upon itself through rotation by an angle of $2\pi/p$. Therefore, the index $p$ somehow determines the number of points of the starburst. In the paper, we only write starburst to economize language, but we note that we refer to star-like $p$-fold symmetric starbursts. An even more surprising fact is that the $p$ index of the $p$-fold symmetry is not always the same, and as a consequence, the number of points of the perceived starburst differs among subjects. This is exemplified even in picture art throughout history, where star representation offers a significant variety of points, though art representation has a strong symbolic character. 

Recently, Rubinstein~\cite{Rubinstein19} has provided a theoretical framework to explain several geometrical properties of starbursts, such as the presence of straight lines passing through the central bright spot. 
Here, we extend Rubinstein's purposes in trying to explain the geometry of starbursts. Specifically, we provide a theoretical explanation of the type of symmetries and the number based on the deep relationship between symmetry, symmetry preservation between wavefronts and caustics, and analysis of the wave aberration's singular points of curvature functions. 
For these purposes, instead of using Cartesian Taylor polynomials as in~\cite{Rubinstein19}, we use Zernike polynomials because they are naturally more suitable for revealing symmetries, and also are the most common mathematical representation of the human eye's wave aberration~\cite{Atchison:23}. Our investigations are based on the singular analysis of the wave aberration function expressed as a combination of Zernike polynomials.  
This procedure allows us to establish the connections between the number and symmetry of some relevant points --the so-called fertile cusps of Gauss~\cite{Barbero:22}-- with the symmetries and the number of starburst points.    

The paper is organized as follows. First, in Section~\ref{sec:theory}, we provide a theoretical framework to explain the symmetries and number of starburst points. For this purpose, we prove a set of propositions concerning symmetries (Section ~\ref{sec:symmetries}) and necessary relations between Zernike coefficients for the appearance of starburst (Section~\ref{sec:conditions}). In the next Section~\ref{sec:numerical}, we provide numerical computations to develop the theoretical results of the previous section. First, we explain how to compute the singular set of the Hessian determinant function and the caustic pattern; second, we compute maps of these quantities for several archetypal examples of starbursts (Section ~\ref{sec:examples}). Finally, in Section~\ref{sec:dis}, we summarize results and discuss their implications.

\section{Theoretical explanation of the number of points and
symmetries of starbursts}\label{sec:theory}
The intersection of caustic surfaces with the retina provides a set (sometimes empty) of points and/or curves.  Caustics are associated with points of the wave aberration where its Hessian determinant (from now on denoted as $G$) is zero (critical set)~\cite{Nye99}. Specifically, caustics are obtained by optically mapping the zero set of $G$; \emph{i.e.}, projecting through ray trajectories pupil plane's points where $G=0$, onto the retina plane. However, the intensity along caustic curves (and vicinity regions) is not uniform. A special mathematical theory developed in the 80's of last century, catastrophe optics~\cite{Berry80, Montaldi:21, Nye99}, predicts that there are some points where there is a hot spot of intensity, which are associated with some points of the wave aberration function called \textit{fertile cusps of Gauss}~\cite{Barbero:22}. They are saddle points, that is, points where the gradient of the Hessian determinant cancels: $\nabla G = 0$~\cite{Barbero:22} and the Hessian matrix for $G$ is indefinite. Also, \textit{fertile} refers to the fact that the saddle cusp of Gauss gives rise to $G$ zero-level contour curves that are close enough to be optically relevant~\cite{Barbero:22}. 

As mentioned before, a necessary condition for a wave aberration function to generate a starburst pattern is that the wavefront concentrates a significant part of the light at the central part of the blurred spot. This is the usual case in most human eyes because wave aberration values are moderately small near the central part of the pupil; indeed, if this were not the case, that eye would have a pathological abnormality. 

But, given the bright central spot, the starburst points are generated due to caustic concentration along them~\cite{Rubinstein19, Xu}. Furthermore, not all the caustics are equally bright; catastrophe theory predicts a much more intense light concentration in the presence of a \textit{cusp} caustic (when the caustic curve contains a discontinuity in the first derivative) or when two \textit{fold} caustics (analytical curves) are close enough.   
Therefore, we conjecture:

\begin{hyp}\label{oh}
Star-like starburst points are perceived only when either a cusp caustic or two very close fold caustics are presented and are separated enough from the central bright spot to be visually resolved. 
\end{hyp}

These two scenarios imply the existence of fertile cusps of Gauss~\cite{Barbero:22} generating two branches of $G$ zero-level contour curves close to it.
When optically mapped onto the retina, these curves may generate a maximum of four-fold and two cusp caustics. 

Therefore, the presence of more than one fertile cusp of Gauss, separated from the center area, is mandatory for generating a starburst pattern. In what follows, we will provide a set of mathematical propositions that set the necessary conditions for the appearance and symmetry of some archetypal starburst patterns. 

\subsection{Symmetry properties of startbursts}\label{sec:symmetries}
 
A strong connection exists between surface symmetries (in $\mathbb{R}^3$) and their curvature properties. For instance, at a non-umbilic point, the maximum reflection symmetry occurs at planes intersecting orthogonally the surface and/or containing a line of curvature~\cite{Bruce:96}. Moreover, a line embedded in a surface, determining a reflectional surface symmetry, usually contains ridge points (points where the lines of curvature are extremum) (\cite[p. 271]{Porteous} and \cite{Bruce:96}). Cusps of Gauss are associated with ridge points, but with the advantage of being easier to compute~\cite{Barbero:22, Porteous}. 
Therefore, computing the cusps of Gauss of the wave aberration provides a way to evaluate symmetries.
An important symmetry-preserving property is given by the following proposition:
\begin{proposition}\label{propA}
The pattern at the retina plane preserves the $p$-fold symmetry of the wave aberration at the pupil plane.
\end{proposition}
\begin{proof}
A $p$-fold symmetry implies the presence of $p$ planes of symmetric reflection.
First, we note that any plane of symmetric reflection of a surface ($S$) intersects it in a geodesic planar curve ($C$). This is because the normal to the surface at any point of a symmetric reflection plane is invariant under reflection, which implies that the normal surface is contained in that plane,
and also the normal to a geodesic curve lies in the normal surface by definition. Additionally, every plane geodesic on a surface is also a line of curvature (\cite[p.~155]{Kreyszig}).

A symmetric reflection plane containing the optical axis is also a meridional plane. Any ray belonging to this meridional plane has zero skewness --part of the vector direction departing from the meridional plane-- because it belongs to a geodesic curve~\cite{Torre:05}, so it remains in that meridional plane through propagation. Therefore, at the retina plane, the curve $C'$ imaged by $C$ is also a planar curve. Furthermore, the focal curve of a line of curvature is itself a geodesic of the focal surface (\cite[p.~176]{Porteous}). As we have shown above, this implies that it also contains a reflectional planar symmetry. Then, the plane of symmetric reflection in the wave aberration and the caustic surface are the same, which intrinsically implies that the $p$-fold symmetry is preserved between the pupil and retina planes.
\end{proof}

The symmetry-preserving property between the wave aberration and the caustic pattern makes it possible, by a simple wave aberration inspection, to realize the $p$-fold symmetry of the starburst pattern and establish a maximum number of starburst points. Indeed, this theoretical result agrees with experimental results obtained using wedge apertures to isolate starburst caustic lines~\cite{Xu}.

Zernike polynomial series commonly represent the wave aberration function of the human eye. Recalling that the angular dependent component of a Zernike polynomial is either $sin(m \theta)$ or $cos(m \theta)$, the $p$-fold symmetry of a Zernike polynomial is given by the so-called azimuthal frequency: $m$. The axial symmetry occurs for terms with $m=0$.
The symmetries are unclear when there is a combination of several Zernike polynomials.
Then, the computation of cusps of Gauss to find geodesic-reflectional symmetric lines emerges as a valuable analytical tool. The number of cusps of Gauss imposes an upper bound on the $p$-fold symmetry and, as a consequence, the number of points of starbursts, as stated by the following proposition:

\begin{proposition}\label{propB} The maximum number of saddle cusps of Gauss is equal to $(n-2)(2n-5)$, being $n$ the degree of the wave aberration function $W$, expressed as a linear combination of Zernike polynomials.
\end{proposition}

\begin{proof} 
If a surface described by a bivariate polynomial has only isolated critical points, then the number of isolated extrema is bounded above. Also, this bound depends upon the degree of the variables of the polynomial~\cite{Qi:04}.
Let's say $d$ is the degree of the bivariate polynomial $f$ containing only isolated critical points. 
Then, the number of saddle points is bounded above by $0.5(d^{2} - d)$ \cite[eq.~(3.1)]{Durfee:93}.

Now, let us consider $W$ the wave aberration function, expressed as a linear combination of Zernike polynomials, being $n$ its degree. 
Thus, the degree of $G=\det({\rm Hess}(W))$ is: $\deg(G) \le 2(n-2)$.
Therefore, the number of saddle points of $G$ is, at most, $(n-2)(2n-5)$.
\end{proof}

With what has been said so far, we analyze which combinations of aberrations give rise to starbursts with different numbers of points. 
The existence of a central intensity core suggests that the wave aberration must contain a combination of axially symmetric terms (defocus plus different orders of spherical aberration) plus some non-axially symmetric terms that break the axial symmetry and generate a $p$-fold symmetry. 

We now study several examples of simple combinations of defocus ($Z_{2}^{0}$), spherical aberration ($Z_{4}^{0}$), and dominant $p$-fold symmetric aberrations ($Z_{n}^{n}$, with $n\ge 3$) given rise to starbursts with several points. All these combinations are admissible as possible human eye aberrations. 

Next, we state that the starburst symmetry pattern is invariant under pupil radius:  

\begin{proposition}\label{propC} The number and normalized radial locations, concerning the pupil size, of saddle cusps of Gauss are invariant to pupil size.
\end{proposition}

\begin{proof} This is a direct consequence of the following fact: Let $f(x,y)$ be a bivariate function and define $g(x,y)=f(\frac xr,\frac yr)$. Then it is clear that $\nabla g(rx,ry)=\frac1r\nabla f(x,y)$ and ${\rm Hess}\, g(rx,ry)=\frac{1}{r^2}{\rm Hess}\,f(x,y)$. Therefore, $(x_0,y_0)$ is a saddle point of $f$ if, and only if, $(rx_0,ry_0)$ is a saddle point of $g$.
\end{proof}

From now on, we are going to consider a wave aberration function expressed as: \begin{equation}\label{eq:W}
W=\alpha Z_2^0+\beta Z_4^0 +\gamma Z_n^n,
\end{equation}
where we use the normalization for Zernike polynomials given in \cite{Laksh2011}. 
We note that the presence of a non-axial dominant $Z_{n}^{n}$ aberration does not imply that $p$-fold symmetry could not appear with other Zernike combinations, but, at least, this is the simplest combination giving rise to that symmetry.

\begin{proposition} Let us consider a wave aberration \eqref{eq:W}. If there is a saddle cusp of Gauss at a distance $\rho>0$ of the center of the pupil, then there are n saddle cusps of Gauss uniformly distributed at the same distance from the center.
\end{proposition}

\begin{proof}
Without loss of generality, we consider $\beta>0$. We perform the partial derivatives of $G$ for $x$ and $y$, and after a change of variable to polar coordinates 
$(x,y)=(\rho \sin\theta, \rho \cos\theta)$,  and taking into account that $\cos(n\theta)=T_{n}(\cos(\theta))$ and  $\sin(n\theta)=\sin(\theta)U_{n-1}(\cos(\theta))$, being $T_n$ and $U_n$ the Chebyshev polynomials of the first and second kind, respectively, we obtain the equations
\begin{eqnarray}
	\rho \sin(\theta) \Big( A(\rho) + B(\rho) U_{n-2}(\cos(\theta)) \Big) = 0, \label{DxAB}
	\\
	\rho \Big( A(\rho) \cos(\theta) - B(\rho) T_{n-1}(\cos(\theta)) \Big) = 0, \label{DyAB}
\end{eqnarray}
where
\[
A(\rho)= 
192 \beta(\sqrt{15} \alpha   -15 \beta) +8640 \beta^2 \rho ^2
-\gamma ^2 (n-2) (n-1)^2 n^2 (n+1) \rho ^{2 n-6}
\]
and
\[
B(\rho)=12 \sqrt{10} \beta  \gamma  (n-1) n^2 \sqrt{n+1} \rho ^{n-2}.
\]
Then, looking at the first equation \eqref{DxAB}, we need to study three different situations:
\begin{enumerate}
\item $\rho=0$.
\item $\rho\ne0$ and $\sin(\theta)=0$.
\item $\rho\ne0$, $\sin(\theta)\ne0$ and $A(\rho) + B(\rho) U_{n-2}(\cos(\theta))=0$.
\end{enumerate}

Let us study each case.

\noindent\textbf{Case  $\rho=0$.}
In this case, the unique critical point is located at the center of the pupil, but it is never a saddle point.

\noindent\textbf{Case $\rho\ne0, \, \sin(\theta)=0$.} This situation, corresponds with either $\theta=0$ or $\theta=\pi$.

Taking $\theta=0$, and assuming  $\rho\ne0$, equation \eqref{DyAB} becomes 
\[
A(\rho) - B(\rho)=0.
\]

Let us suppose that $(\rho^*,0)$ is a solution of \eqref{DxAB} and \eqref{DyAB} with $\rho^*\ne0$. 
Then, for $(\rho^*,\frac{2k\pi}{n})$, $k=1,\dots, n-1$, equations  \eqref{DxAB} and \eqref{DyAB} simplify to
\[
\begin{aligned}
\sin\Big(\frac{2k\pi}{n}\Big) \Big( A(\rho^*)+B(\rho^*) U_{n-2}\Big(\cos\Big(\frac{2k\pi}{n}\Big)\Big) \Big)&=0,
\\
 A(\rho^*)\cos\Big(\frac{2k\pi}{n}\Big)-B(\rho^*) T_{n-1}\Big(\cos\Big(\frac{2k\pi}{n}\Big)\Big) &=0.
\end{aligned}
\]

In the case $k\ne n/2$, the equations reduce to
\[
\begin{aligned}
	 A(\rho^*)-B(\rho^*) &=0,
	\\
\big(	A(\rho^*)-B(\rho^*) \big) \cos\big(\frac{2k\pi}{n}\big) &=0,
\end{aligned}
\]
which is held by hypothesis. 

In the case $k=n/2$, the second equation is the same as before, and the first one vanishes since $\sin\big(\frac{2k\pi}{n}\big)=0$.

With this, we have proved that if $(\rho^*,0)$ is a critical point, then $(\rho^*,\frac{2k\pi}{n})$ for $k=0,...,n-1$ are also critical points. 

For the case  $\theta=\pi$, equations for computing the critical points reduce to \[A(\rho )+(-1)^{n-1} B(\rho )=0.\]
An analog argument, as in the previous case, leads us to the same conclusion: if $(\rho^*,\pi)$ is critical point, then $(\rho^*,\pi+\frac{2k\pi}{n})$ for $k=0,...,n-1$ are also critical points. 

\noindent\textbf{Case $\rho\ne0, \, \sin(\theta)\ne0$.}
Now, equations for critical points are:
\begin{eqnarray*}
 A(\rho) + B(\rho) U_{n-2}(\cos(\theta)) = 0,
	\\
 A(\rho) \cos(\theta) - B(\rho) T_{n-1}(\cos(\theta)) = 0.
\end{eqnarray*}
Then, taking apart $A(\rho)$ in the first equation and substituting it in the second, one gets
\begin{eqnarray*}
	A(\rho) =- B(\rho) U_{n-2}(\cos(\theta)),
	\\
	B(\rho)\big(  U_{n-2}(\cos(\theta)) \cos(\theta) +  T_{n-1}(\cos(\theta))\big) = 0.
\end{eqnarray*}
From these, considering properties of Chebyshev polynomials, we get
$$
U_{n-2}(\cos\theta) =-\cos(n\theta)=\pm1.
$$
The sign of the last equality depends on whether $n\theta=2k\pi$ or $n\theta=(2k+1)\pi$, with $k\in\mathbb{Z}$.
 
With this, we finally have $\sin(n\theta)=0$ and
\begin{equation}
A(\rho) - B(\rho)= 0 \text{ if } n\theta=2k\pi,
\end{equation}
or 
\begin{equation}
A(\rho) + B(\rho)= 0 \text{ if } n\theta=(2k+1)\pi.
\end{equation}

Moreover, one can prove that when $\theta=(2k+1)\pi/n$, $k=0,\dots,n-1$, then the Hessian determinant of $G$ is always the same; therefore, all the corresponding critical points are saddle points or not. The same situation occurs in the case $\theta=2k\pi/n$. 
\end{proof}

\begin{proposition}\label{th:fertile} Let us consider a wave aberration function defined as in \eqref{eq:W}, with $n \in \{3, 4, 5, 6\}$. Then, the number $s$ of saddle cusps of Gauss inside the pupil is:
\begin{itemize}
\item[$\circ$] $s=0$ or $s=n$ when $n=3,4$,
\item[$\circ$] $s=0$, $s=n$ or $s=2n$ when $n=5,6$. 
\end{itemize}
Moreover, if $s=n$, then the points are uniformly distributed in a circle centered at the origin. Also, if $s=2n$, they are uniformly distributed in two circles of different radius, $n$ points in each. 
\end{proposition}

\begin{proof} We distinguish for different values of $n$.

\noindent\textbf{Case $n=3$.} 
In this case, both equations $A(\rho) \pm B(\rho)= 0$ simplify to two quadratic equations in $\rho$:
\[
180 \beta ^2 \rho ^2 \pm 9 \sqrt{10} \beta  \gamma  \rho + 4 \beta  \Big(\sqrt{15} \alpha -15 \beta \Big)-3 \gamma ^2 =0,
\]
which we can solve explicitly and obtain the four solutions
\[
\rho = \frac{\pm 3 \gamma  \pm \sqrt{32 \beta  \left(15 \beta -\sqrt{15} \alpha \right)+33 \gamma ^2}}{12 \sqrt{10} \beta }.
\]
Observe here that, if they are real, there are two positive values and two negative ones.
To give conditions for these values to be real, $0<\rho<1$, and the corresponding critical points to be saddle cusps of Gauss, we need to define the following parameters:
\begin{equation}\label{eq:alphasn=3}
\begin{aligned}
&\alpha_1^\pm=\frac{-120 \beta ^2\pm9 \sqrt{10} \beta  \gamma +3 \gamma ^2}{4 \sqrt{15} \beta}, \quad 
\alpha_2=\frac{60 \beta ^2+3 \gamma ^2}{4 \sqrt{15} \beta},\\ 
&\alpha_3=\frac{480 \beta ^2+33 \gamma ^2}{32 \sqrt{15} \beta}.
\end{aligned}
\end{equation}

\begin{figure}[ht]
\centering
	\begin{tikzpicture}[xscale=0.25,yscale=0.04]
		\draw[->] (-20,0) -- (22,0) node[below right] {$\gamma$};
		\draw[->] (0,-15) -- (0,120) node[above left] {$\alpha$};
%% gamma>0
% a3
\path[domain=0:20,color=red, draw, name path=a3] plot(\x,{(480+33*(\x)^2)/(32*sqrt(15))}) node[above right] {$\alpha_3$};
\path[domain=0:{4*sqrt(10)},color=red, draw, name path=a3first] plot(\x,{(480+33*(\x)^2)/(32*sqrt(15))});
\path[domain={4*sqrt(10)}:20,color=red, draw, name path=a3second] plot(\x,{(480+33*(\x)^2)/(32*sqrt(15))});
% a1p
\path[domain=0:20,color=blue, draw, name path=a1p] plot(\x,{(-120+9*sqrt(10)*\x+3*(\x)^2)/(4*sqrt(15))})  node[right] {$\alpha_1^+$};
\path[domain=0:{4*sqrt(10)},color=blue, draw, name path=a1pfirst] plot(\x,{(-120+9*sqrt(10)*\x+3*(\x)^2)/(4*sqrt(15))});
\path[domain={4*sqrt(10)}:20,color=blue, draw, name path=a1psecond] plot(\x,{(-120+9*sqrt(10)*\x+3*(\x)^2)/(4*sqrt(15))});
% a2
\path[domain=0:20,color=orange, draw, name path=a2] plot(\x,{(60+3*(\x)^2)/(4*sqrt(15))}) node[right] {$\alpha_2$};
\path[domain=0:{4*sqrt(10)},color=orange, draw, name path=a2first] plot(\x,{(60+3*(\x)^2)/(4*sqrt(15))});
\path[domain={4*sqrt(10)}:20,color=orange, draw, name path=a2second] plot(\x,{(60+3*(\x)^2)/(4*sqrt(15))});
% a1m
\path[domain=0:20,color=olive, draw, name path=a1m] plot(\x,{(-120-9*sqrt(10)*\x+3*(\x)^2)/(4*sqrt(15))}) node[right] {$\alpha_1^-$};
\path[domain=0:{4*sqrt(10)},color=olive, draw, name path=a1mfirst] plot(\x,{(-120-9*sqrt(10)*\x+3*(\x)^2)/(4*sqrt(15))}) ;
\path[domain={4*sqrt(10)}:20,color=olive, draw, name path=a1msecond] plot(\x,{(-120-9*sqrt(10)*\x+3*(\x)^2)/(4*sqrt(15))}) ;
\tikzfillbetween[of=a1m and a2, on layer=bl]{blue, opacity=0.1};
\tikzfillbetween[of=a2first and a3first,on layer=or]{orange, opacity=0.1};
\tikzfillbetween[of=a2second and a1psecond,on layer=or]{orange, opacity=0.1};

%% gamma<0
% a3
\path[domain=0:-20,color=red, draw, name path=a3] plot(\x,{(480+33*(\x)^2)/(32*sqrt(15))}) node[above left] {$\alpha_3$};
\path[domain=0:{-4*sqrt(10)},color=red, draw, name path=a3first] plot(\x,{(480+33*(\x)^2)/(32*sqrt(15))});
\path[domain={-4*sqrt(10)}:-20,color=red, draw, name path=a3second] plot(\x,{(480+33*(\x)^2)/(32*sqrt(15))});
% a1p
\path[domain=0:-20,color=blue, draw, name path=a1p] plot(\x,{(-120+9*sqrt(10)*(\x)+3*(\x)^2)/(4*sqrt(15))})  node[left] {$\alpha_1^+$};
\path[domain=0:{-4*sqrt(10)},color=blue, draw, name path=a1pfirst] plot(\x,{(-120+9*sqrt(10)*\x+3*(\x)^2)/(4*sqrt(15))});
\path[domain={-4*sqrt(10)}:-20,color=blue, draw, name path=a1psecond] plot(\x,{(-120+9*sqrt(10)*\x+3*(\x)^2)/(4*sqrt(15))});
% a2
\path[domain=0:-20,color=orange, draw, name path=a2] plot(\x,{(60+3*(\x)^2)/(4*sqrt(15))}) node[left] {$\alpha_2$};
\path[domain=0:{-4*sqrt(10)},color=orange, draw, name path=a2first] plot(\x,{(60+3*(\x)^2)/(4*sqrt(15))});
\path[domain={-4*sqrt(10)}:-20,color=orange, draw, name path=a2second] plot(\x,{(60+3*(\x)^2)/(4*sqrt(15))});
% a1m
\path[domain=0:-20,color=olive, draw, name path=a1m] plot(\x,{(-120-9*sqrt(10)*(\x)+3*(\x)^2)/(4*sqrt(15))}) node[left] {$\alpha_1^-$};
\path[domain=0:{-4*sqrt(10)},color=olive, draw, name path=a1mfirst] plot(\x,{(-120-9*sqrt(10)*\x+3*(\x)^2)/(4*sqrt(15))}) ;
\path[domain={-4*sqrt(10)}:-20,color=olive, draw, name path=a1msecond] plot(\x,{(-120-9*sqrt(10)*\x+3*(\x)^2)/(4*sqrt(15))}) ;
\tikzfillbetween[of=a1p and a2, on layer=or]{orange, opacity=0.1};
\tikzfillbetween[of=a2first and a3first, on layer=bl]{blue, opacity=0.1};
\tikzfillbetween[of=a2second and a1msecond, on layer=bl]{blue, opacity=0.1};

		\draw[-,very thin, dotted ] ({4*sqrt(10)},{12*sqrt(15)}) -- ({4*sqrt(10)},0) node[below=2pt] {\tiny$4\sqrt{10}\beta$};

		\draw[-,thick] ({4*sqrt(10)},1) -- ({4*sqrt(10)},-1);

		\draw[-,very thin, dotted ] ({-4*sqrt(10)},{12*sqrt(15)}) -- ({-4*sqrt(10)},0) node[below=2pt] {\tiny$-4\sqrt{10}\beta$};

		\draw[-,thick] ({-4*sqrt(10)},1) -- ({-4*sqrt(10)},-1);
		
% tiks en Y
		\draw[-,very thin, dotted ] (0,{-2*sqrt(15)}) -- (0,{-2*sqrt(15)}) node[below left=0pt] {\tiny$-2\sqrt{15}\beta$};
		\draw[-,very thin, dotted ] (0,{sqrt(15)}) -- (0,{sqrt(15)}) node[above left=0pt] {\tiny$\sqrt{15}\beta$};

		\draw[-,very thin, dotted ] ({4*sqrt(10)},{12*sqrt(15)}) -- (0,{12*sqrt(15)}) node[left=2pt] {\tiny$12\sqrt{15}\beta$};

		\draw[-,very thin, dotted ] ({-4*sqrt(10)},{12*sqrt(15)}) -- (0,{12*sqrt(15)});
		\draw[-,thick] (0.2,{-2*sqrt(15)}) -- (-0.2,{-2*sqrt(15)});
		\draw[-,thick] (0.2,{sqrt(15)}) -- (-0.2,{sqrt(15)});

		\draw[-,thick] (0.2,{12*sqrt(15)}) -- (-0.2,{12*sqrt(15)});

% ejemplo numerico
\node at (1,0) (ex) {$\bullet$}; 
	\end{tikzpicture}
\caption{Regions of values for $\alpha$ and $\gamma$ (depending on $\beta$) providing saddle cusps of Gauss inside the pupil (case $n=3$). Blue areas correspond with saddle points with angular coordinates $\theta=\pi/3,\pi,5\pi/3$. Orange areas correspond with saddle points with angular coordinates $\theta=0,2\pi/3,4\pi/3$. The black bullet corresponds with values of example \eqref{eq:3star}. Expressions for $\alpha_1^\pm,\alpha_2,\alpha_3$ are given in \eqref{eq:alphasn=3}.} 
\label{fig:regn3}
\end{figure}
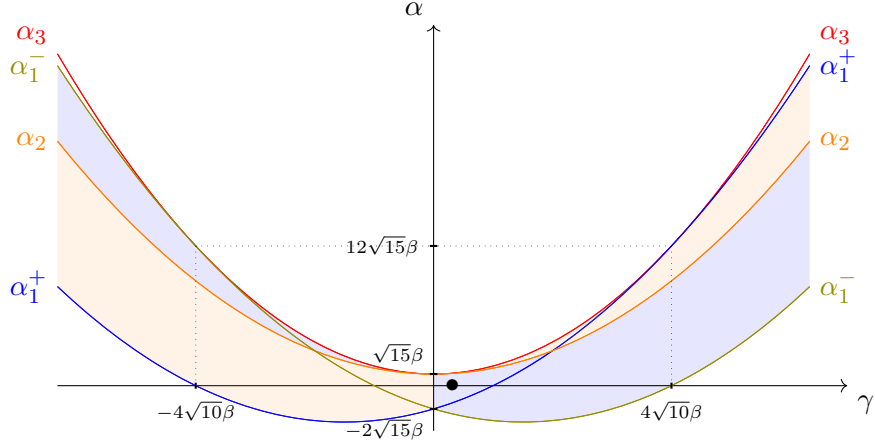

\noindent 
After some calculations (done using {\em Mathematica}) we have the following conditions (see Figure~\ref{fig:regn3}):
\begin{itemize}
\item there are saddle cusps of Gauss inside the pupil, $(\rho,\theta)$ with angular values $\theta=0,2\pi/3,4\pi/3$, when
\smallskip

\centerline{\renewcommand{\arraystretch}{1.5}
\begin{tabular}{|>{\centering\arraybackslash}p{3.5cm}|>{\centering\arraybackslash}p{3.5cm}|>{\centering\arraybackslash}p{3.5cm}|}
\hline $\gamma<0$ \par $\alpha_1^+<\alpha<\alpha_2$
 &
 $0<\gamma<4\sqrt{10}\beta$ \par $\alpha_2<\alpha<\alpha_3$
 &
 $\gamma>4\sqrt{10}\beta$ \par $\alpha_2<\alpha<\alpha_1^+$
 \\\hline
\end{tabular}}\smallskip

\item there are saddle cusps of Gauss inside the pupil, $(\rho,\theta)$ with angular values $\theta=\pi/3,\pi,5\pi/3$, when
\smallskip

\centerline{\renewcommand{\arraystretch}{1.5}
\begin{tabular}{|>{\centering\arraybackslash}p{3.5cm}|>{\centering\arraybackslash}p{3.5cm}|>{\centering\arraybackslash}p{3.5cm}|}
\hline 
$\gamma>0$ \par $\alpha_1^-<\alpha<\alpha_2$
 &
 $-4\sqrt{10}\beta<\gamma<0$ \par $\alpha_2<\alpha<\alpha_3$
 &
 $\gamma<-4\sqrt{10}\beta$ \par $\alpha_2<\alpha<\alpha_1^-$
 \\\hline
\end{tabular}}\smallskip
\end{itemize}
%%%%%%%%%%%%%%%5
%%%%%%%%%%%%%%%%%%

\noindent\textbf{Case $n=4$.}
In this case, we proceed similarly to the previous one. Equations $A(\rho) \pm B(\rho)=0$ become

\begin{equation}\label{ab4}
	\begin{aligned}
		2 \beta(\sqrt{15} \alpha  -15 \beta)  +15 \rho^2(6 \beta^2-\gamma^2 		\pm 2 \sqrt{2} \beta  \gamma  )	
=0,
\end{aligned}
\end{equation}
which we can easily solve explicitly, obtaining two positive solutions, one from each equation:	  
\begin{equation}\label{rho1}
	\rho=\sqrt{\frac{2}{15}}\sqrt{\frac{\beta(15\beta-\sqrt{15}\alpha)}{6\beta^2\pm2\sqrt{2}\beta\gamma -\gamma^2}}.
\end{equation}
We observe that the solution with the minus sign corresponds with the equation $A(\rho) - B(\rho)=0$ which comes with the angles $\theta=0,\pi/2,\pi,3\pi/2$, whereas the solution with the plus sign corresponds with  $A(\rho) + B(\rho)=0$ which comes with the angles $\theta=\pi/4,3\pi/4,5\pi/4,7\pi/4$.

Again using {\em Mathematica}, we explore the conditions to guarantee the existence of saddle cusps of Gauss inside the pupil. Let us define:
\begin{equation}\label{eq:alphasn=4}
\alpha_1^\pm=\frac{-60\beta^2\pm30\sqrt{2} \beta\gamma +15\gamma^2}{2\sqrt{15}\beta}.
\end{equation}

	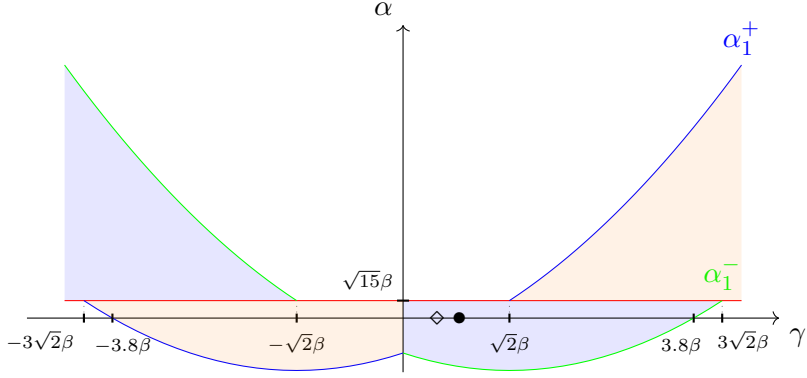
\begin{figure}[ht]
\centering	\begin{tikzpicture}[xscale=1,yscale=0.06]
	\draw[->] (-5,0) -- (5,0) node[below right] {$\gamma$};
	\draw[->] (0,-12) -- (0,65) node[above left] {$\alpha$};
%% gamma>0

% sqrt(15)
	\path[domain=0:4.5,color=red, draw, name path=sqrt15p] plot(\x,{sqrt(15)});
	\path[domain=-4.5:0,color=red, draw, name path=sqrt15m] plot(\x,{sqrt(15)});

% a1p = {0.5*sqrt(15)*(-4+2*sqrt(2)*(\x)+(\x)^2)}
	\path[domain=sqrt(2):4.5,color=blue, draw, name path=a1p] plot(\x,{0.5*sqrt(15)*(-4+2*sqrt(2)*(\x)+(\x)^2)})  node[above] {$\alpha_1^+$};
	\path[domain=-3*sqrt(2):0,color=blue, draw, name path=a1pm] plot(\x,{0.5*sqrt(15)*(-4+2*sqrt(2)*(\x)+(\x)^2)}) ;% node[above] {$\alpha_1^+$};

% a1m = {0.5*sqrt(15)*(-4-2*sqrt(2)*(\x)+(\x)^2)}
	\path[domain=0:3*sqrt(2),color=green, draw, name path=a1m] plot(\x,{0.5*sqrt(15)*(-4-2*sqrt(2)*(\x)+(\x)^2)}) node[above] {$\alpha_1^-$};
	\path[domain=-4.5:-sqrt(2),color=green, draw, name path=a1mm] plot(\x,{0.5*sqrt(15)*(-4-2*sqrt(2)*(\x)+(\x)^2)});

% rellenos
	\tikzfillbetween[of=a1m and sqrt15p, on layer=bl]{blue, opacity=0.1};
	\tikzfillbetween[of=a1mm and sqrt15m, on layer=bl]{blue, opacity=0.1};
	\tikzfillbetween[of=a1p and sqrt15p, on layer=or]{orange, opacity=0.1};
	\tikzfillbetween[of=a1pm and sqrt15m, on layer=or]{orange, opacity=0.1};

% tiks en X
		\draw[-,very thin, dotted ] ({sqrt(2)},0) -- ({sqrt(2)},0) node[below=2pt] {\tiny$\sqrt{2}\beta$};
		\draw[-,very thin, dotted ] ({sqrt(2)},0) -- ({sqrt(2)},{sqrt(15)});
		\draw[-,thick] ({sqrt(2)},1) -- ({sqrt(2)},-1);

		\draw[-,very thin, dotted ] ({3*sqrt(2)},0) -- ({3*sqrt(2)},0) node[below=1pt] {\tiny\qquad$3\sqrt{2}\beta$};
		\draw[-,very thin, dotted ] ({3*sqrt(2)},0) -- ({3*sqrt(2)},{sqrt(15)});
		\draw[-,thick] ({3*sqrt(2)},1) -- ({3*sqrt(2)},-1);

		\draw[-,very thin, dotted ] ({-sqrt(2)},0) -- ({-sqrt(2)},0) node[below=2pt] {\tiny$-\sqrt{2}\beta$};
		\draw[-,very thin, dotted ] ({-sqrt(2)},0) -- ({-sqrt(2)},{sqrt(15)});
		\draw[-,thick] ({-sqrt(2)},1) -- ({-sqrt(2)},-1);

		\draw[-,very thin, dotted ] ({-3*sqrt(2)},0) -- ({-3*sqrt(2)},0) node[below=1pt] {\tiny$-3\sqrt{2}\beta$\qquad\qquad\,};
		\draw[-,very thin, dotted ] ({-3*sqrt(2)},0) -- ({-3*sqrt(2)},{sqrt(15)});
        \draw[-,thick] ({-3*sqrt(2)},1) -- ({-3*sqrt(2)},-1);

		\draw[-,very thin, dotted ] ({-sqrt(2)-sqrt(6)},0) -- ({-sqrt(2)-sqrt(6)},0) node[below=4pt] {\tiny\quad$-3.8\beta$};
		\draw[-,thick] ({-sqrt(2)-sqrt(6)},1) -- ({-sqrt(2)-sqrt(6)},-1);

		\draw[-,very thin, dotted ] ({sqrt(2)+sqrt(6)},0) -- ({sqrt(2)+sqrt(6)},0) node[below=4pt] {\tiny$3.8\beta$\quad\,};
		\draw[-,thick] ({sqrt(2)+sqrt(6)},1) -- ({sqrt(2)+sqrt(6)},-1);

% tiks en Y
		\draw[-,very thin, dotted ] (0,{sqrt(15)}) -- (0,{sqrt(15)}) node[above left=0pt] {\tiny$\sqrt{15}\beta$};
		\draw[-,thick] (0.08,{sqrt(15)}) -- (-0.08,{sqrt(15)});

% ejemplo numerico
\node at (0.75,0) (ex) {$\bullet$}; 
\node at (0.45,0) (ex1) {$\diamond$}; 
	\end{tikzpicture}
\caption{Regions of values for $\alpha$ and $\gamma$ (depending on $\beta$) providing saddle cusps of Gauss inside the pupil (case $n=4$). Blue areas correspond with saddle points with  $\theta=\pi/4,3\pi/4,5\pi/4,7\pi/4$. Orange areas correspond with saddle points with angular coordinates $\theta=0,\pi/2,\pi,3\pi/2$. Black bullet corresponds with election of values for example \eqref{eq:4star} and diamond with example \eqref{eq:8stars} Expressions for $\alpha_1^\pm$ are given in \eqref{eq:alphasn=4}}
	\label{fig:regn4}
\end{figure}

\noindent
Then, there are saddle cusps of Gauss inside the pupil, $(\rho,\theta)$ (see Figure~\ref{fig:regn4}):
\begin{itemize}%\itemindent35pt
\item  with $\theta=0,\pi/2,\pi,3\pi/2$, when
\smallskip

\centerline{\renewcommand{\arraystretch}{1.5}
\begin{tabular}{|>{\centering\arraybackslash}p{4cm}|>{\centering\arraybackslash}p{4cm}|}
\hline 
$-3\sqrt{2} \beta<\gamma<0$ \par $\alpha_1^+<\alpha<\displaystyle\sqrt{15}\beta$
 &
 $\gamma>\sqrt{2} \beta$ \par $\displaystyle\sqrt{15}\beta<\alpha<\alpha_1^+$
 \\\hline
\end{tabular}}\smallskip

\item with $\theta=\pi/4,3\pi/4,5\pi/4,7\pi/4$ when
\smallskip

\centerline{\renewcommand{\arraystretch}{1.5}
\begin{tabular}{|>{\centering\arraybackslash}p{4cm}|>{\centering\arraybackslash}p{4cm}|}
\hline 
$0<\gamma<3\sqrt{2} \beta$ \par $\alpha_1^-<\alpha<\displaystyle\sqrt{15}\beta$
 &
 $\gamma<-\sqrt{2} \beta$ \par $\displaystyle\sqrt{15}\beta<\alpha<\alpha_1^-$
 \\\hline
\end{tabular}}\smallskip

%\begin{itemize}\itemindent25pt
%	\item[$\circ$] $0<\gamma<3\sqrt{2} \beta$ and $\alpha_1^-<\alpha<\displaystyle\sqrt{15}\beta$;
%	\item[$\circ$] $\gamma<-\sqrt{2} \beta$ and $\displaystyle\sqrt{15}\beta<\alpha<\alpha_1^-$.	
%\end{itemize}
\end{itemize}

\noindent\textbf{Case $n=5$.} In this case, we proceed again as previously. Equations $A(\rho) \pm B(\rho)=0$ become
\[
2 \sqrt{15} \beta(\alpha  -\sqrt{15} \beta )
 +90 \beta ^2 \rho ^2\pm25 \sqrt{15} \beta  \gamma  \rho ^3-75 \gamma ^2 \rho ^4 = 0,
\]
whose solutions can be explicitly solved with {\emph Mathematica}.
In the discussion, we will need some parameters that we define here
\begin{equation}\label{eq:alphasn=5}
\begin{aligned}
&\alpha_1^\pm=\frac{-60\beta^2\pm25\sqrt{15} \beta\gamma +75\gamma^2}{2\sqrt{15}\beta} , \qquad
\alpha_2^\pm=\frac{9 \left(4561\pm445 \sqrt{89}\right)\beta^3}{1024 \sqrt{15}\gamma ^2},
\\
&\gamma_1^\pm = \frac{\sqrt{15} \left(\sqrt{89}\pm5\right)\beta}{40} .
\end{aligned}
\end{equation}

%%%%%%%%%
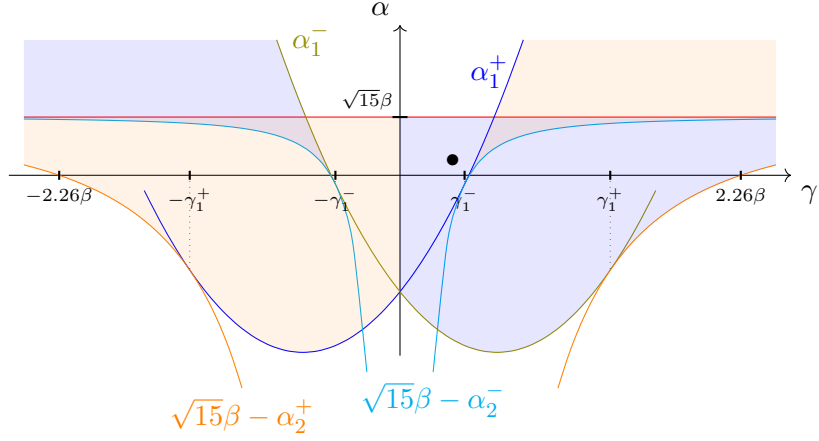
\begin{figure}[ht]
\centering
\begin{tikzpicture}[xscale=2,yscale=0.2]
% los ejes
	\draw[->] (-2.6,0) -- (2.6,0) node[below right] {$\gamma$};
	\draw[->] (0,-12) -- (0,10) node[above left] {$\alpha$};

% Todas las funciones, sobre todo el dominio, encima para que se vean las gráficas
% sqrt(15)
	\path[domain=-2.5:2.5,color=red, draw] plot[smooth] (\x,{sqrt(15)});

% a2p = 0.5*(-4*sqrt(15) + 25*(\x) + 5*sqrt(15)*(\x)^2)
	\path[domain=-1.7:0.82,color=blue, draw] plot[smooth] (\x,{0.5*(-4*sqrt(15) + 25*(\x) + 5*sqrt(15)*(\x)^2)})  node[below left=2pt] {$\alpha_1^+$};

% a2m = 0.5*(-4*sqrt(15) - 25*(\x) + 5*sqrt(15)*(\x)^2)
	\path[domain=1.7:-0.82,color=olive, draw] plot[smooth] (\x,{0.5*(-4*sqrt(15) - 25*(\x) + 5*sqrt(15)*(\x)^2)})  node[right=2pt] {$\alpha_1^-$};

% sqrt(15)-a1p = sqrt(15)-19.8773/(\x)^2
\path[domain=-2.5:-1.05,color=orange, draw] plot[smooth] (\x,{sqrt(15)-19.8773*(\x)^(-2)})  node[below] {$\sqrt{15}\beta-\alpha_2^+$};
\path[domain=1.05:2.5,color=orange, draw] plot[smooth] (\x,{sqrt(15)-19.8773*(\x)^(-2)});

% sqrt(15)-a1m = sqrt(15)-0.823489/(\x)^2
\path[domain=-2.5:-0.22,color=cyan, draw] plot[smooth] (\x,{sqrt(15)-0.823489*(\x)^(-2)});
\path[domain=2.5:0.22,color=cyan, draw] plot[smooth] (\x,{sqrt(15)-0.823489*(\x)^(-2)}) node[below] {$\sqrt{15}\beta-\alpha_2^-$};
	\clip (-3,-12) rectangle (3,9);
% funciones por trozos para hacer el relleno (esto no sé cómo usarlo pero ahí tenemos los valores)
\coordinate (g1m) at (0.429318,0) ;
\coordinate (g2m) at (0.625985,0);
\coordinate (g1p) at (1.39756,0);
\coordinate (g2p) at (1.91698,0);

% sqrt(15)
	\path[domain=-2.5:-1.39756, name path=2sqrt15] plot[smooth](\x,{sqrt(15)});
	\path[domain=-1.39756:0, name path=3sqrt15] plot[smooth](\x,{sqrt(15)});
	\path[domain=0:1.39756, name path=4sqrt15] plot[smooth](\x,{sqrt(15)});
	\path[domain=1.39756:2.5, name path=5sqrt15] plot[smooth](\x,{sqrt(15)});

% a2p = 0.5*(-4*sqrt(15) + 25*(\x) + 5*sqrt(15)*(\x)^2)
	\path[domain=-1.39756:0, name path=3a2p] plot[smooth](\x,{0.5*(-4*sqrt(15) + 25*(\x) + 5*sqrt(15)*(\x)^2)});
	\path[domain=0.429318:2.5, name path=6a2p] plot[smooth](\x,{0.5*(-4*sqrt(15) + 25*(\x) + 5*sqrt(15)*(\x)^2)});

% a2m = 0.5*(-4*sqrt(15) - 25*(\x) + 5*sqrt(15)*(\x)^2)
	\path[domain=-2.5:-0.429318, name path=1a2m] plot[smooth] (\x,{0.5*(-4*sqrt(15) - 25*(\x) + 5*sqrt(15)*(\x)^2)});
	\path[domain=0:1.39756, name path=4a2m] plot[smooth] (\x,{0.5*(-4*sqrt(15) - 25*(\x) + 5*sqrt(15)*(\x)^2)});

% sqrt(15)-a1p = sqrt(15)-19.8773/(\x)^2
	\path[domain=-2.5:-1.39756, name path=2a1p] plot[smooth] (\x,{sqrt(15)-19.8773*(\x)^(-2)});
	\path[domain=1.39756:2.5, name path=5a1p] plot[smooth] (\x,{sqrt(15)-19.8773*(\x)^(-2)});
	
% sqrt(15)-a1m = sqrt(15)-0.823489/(\x)^2
	\path[domain=-2.5:-0.429318, name path=1a1m] plot[smooth] (\x,{sqrt(15)-0.823489*(\x)^(-2)});
	\path[domain=0.429318:2.5, name path=6a1m] plot[smooth] (\x,{sqrt(15)-0.823489*(\x)^(-2)});

% rellenos
	\tikzfillbetween[of=1a2m and 1a1m, on layer = main]{blue, opacity=0.1};
	\tikzfillbetween[of=2a1p and 2sqrt15, on layer = main]{orange, opacity=0.1};
	\tikzfillbetween[of=3sqrt15 and 3a2p, on layer = main]{orange, opacity=0.1};
	\tikzfillbetween[of=4a2m and 4sqrt15, on layer = main]{blue, opacity=0.1};
	\tikzfillbetween[of=5a1p and 5sqrt15, on layer = main]{blue, opacity=0.1};
	\tikzfillbetween[of=6a1m and 6a2p, on layer = main]{orange, opacity=0.1};

% tiks en X

		\draw[-,very thin, dotted ] ({-1.39756},0) -- ({-1.39756},0) node[below=1pt] {\tiny$-\gamma_1^+$};
		\draw[-,very thin, dotted ] ({-1.39756},0) -- ({-1.39756},{-6.30388});
		\draw[-,thick] ({-1.39756},0.3) -- ({-1.39756},-0.3);
			
		\draw[-,very thin, dotted ] ({1.39756},0) -- ({1.39756},0) node[below=1pt] {\tiny$\gamma_1^+$};
		\draw[-,very thin, dotted ] ({1.39756},0) -- ({1.39756},{-6.30388});
		\draw[-,thick] ({1.39756},0.3) -- ({1.39756},-0.3);

		\draw[-,very thin, dotted ] ({-0.429318},0) -- ({-0.429318},0) node[below=1pt] {\tiny$-\gamma_1^-$};
		\draw[-,very thin, dotted ] ({-0.429318},0) -- ({-0.429318},{-0.594872});
		\draw[-,thick] ({-0.429318},0.3) -- ({-0.429318},-0.3);

		\draw[-,very thin, dotted ] ({0.429318},0) -- ({0.429318},0) node[below=1pt] {\tiny$\gamma_1^-$};
		\draw[-,very thin, dotted ] ({0.429318},0) -- ({0.429318},{-0.594872});
		\draw[-,thick] ({0.429318},0.3) -- ({0.429318},-0.3);

		\draw[-,very thin, dotted ] ({2.26546},0) -- ({2.26546},0) node[below=1pt] {\tiny$2.26\beta$};

		\draw[-,thick] ({2.26546},0.3) -- ({2.26546},-0.3);

		\draw[-,very thin, dotted ] ({-2.26546},0) -- ({-2.26546},0) node[below=1pt] {\tiny$-2.26\beta$};

		\draw[-,thick] ({-2.26546},0.3) -- ({-2.26546},-0.3);

% tiks en Y
			\draw[-,very thin, dotted ] (0,{sqrt(15)}) -- (0,{sqrt(15)}) node[above left=-1pt] {\tiny$\sqrt{15}\beta$};
			\draw[-,thick] (0.05,{sqrt(15)}) -- (-0.05,{sqrt(15)});

% ejemplo numerico
\node at (0.35,1) (ex) {$\bullet$}; 
    \end{tikzpicture}
\caption{Regions of values for $\alpha$ and $\gamma$ (depending on $\beta$) providing saddle cusps of Gauss inside the pupil (case $n=5$). Blue areas correspond with saddle points with angular coordinates $\theta=\pi/5,3\pi/5,\pi,7\pi/5,9\pi/5$. Orange areas correspond with saddle points with $\theta=0,2\pi/5,4\pi/5,6\pi/5,8\pi/5$. The black bullet corresponds with values of example \eqref{eq:5star}. Expressions for $\alpha_1^\pm, \alpha_2^\pm, \gamma_1^\pm$ are given in \eqref{eq:alphasn=5}}
\label{fig:regn5}
\end{figure}

\noindent
Then, there are saddle cusps of Gauss inside the pupil, $(\rho,\theta)$ in the following conditions (see Figure~\ref{fig:regn5}):
\begin{itemize}
\item with $\theta=\frac{\pi}{5},\frac{3\pi}{5},\pi,\frac{7\pi}{5},\frac{9\pi}{5}$, if\smallskip

\centerline{\renewcommand{\arraystretch}{1.5}
\begin{tabular}{|>{\centering\arraybackslash}p{4cm}|>{\centering\arraybackslash}p{3.2cm}|>{\centering\arraybackslash}p{4cm}|}
\hline 
 $\gamma<-\gamma_1^-$ \par $\sqrt{15}\beta-\alpha_2^-<\alpha<\alpha_1^+$
 &
 $0<\gamma<\gamma_1^+$  \par $\alpha_1^-<\alpha<\sqrt{15}\beta$
 &
 $\gamma>\gamma_1^+$  \par $\sqrt{15}\beta-\alpha_2^+<\alpha<\sqrt{15}\beta$
 \\\hline
\end{tabular}}\smallskip

\item with $\theta=0,\frac{2\pi}{5},\frac{4\pi}{5},\frac{6\pi}{5},\frac{8\pi}{5}$, if \smallskip

\centerline{\renewcommand{\arraystretch}{1.5}
\begin{tabular}{|>{\centering\arraybackslash}p{4cm}|>{\centering\arraybackslash}p{3.2cm}|>{\centering\arraybackslash}p{4cm}|}
\hline 
 $\gamma<-\gamma_1^+$ \par $\sqrt{15}\beta-\alpha_2^+<\alpha<\sqrt{15}\beta$
 &
 $-\gamma_1^+<\gamma<0$ \par $\alpha_1^+<\alpha<\sqrt{15}\beta$
 &
 $\gamma>\gamma_1^-$ \par $\sqrt{15}\beta-\alpha_2^-<\alpha<\alpha_1^+$
 \\\hline
\end{tabular}}\smallskip
\end{itemize}

In Figure~\ref{fig:regn5} we can see the different regions. We observe that there are two pieces with saddle cusps of Gauss located at two radii with different angles, namely:\smallskip

\centerline{\renewcommand{\arraystretch}{1.5}
\begin{tabular}{|>{\centering\arraybackslash}p{5.8cm}|>{\centering\arraybackslash}p{5.8cm}|}
\hline 
$\gamma>\gamma_1^-$ \par $\sqrt{15}\beta-\alpha_2^-<\alpha<\min\{\alpha_1^+,\sqrt{15}\beta\}$
 &
 $\gamma<-\gamma_1^-$ \par $\sqrt{15}\beta-\alpha_2^-<\alpha<\min\{\alpha_1^-,\sqrt{15}\beta\}$
 \\\hline
\end{tabular}}\medskip

\noindent\textbf{Case $n=6$.} In this case, equations $A(\rho) \pm B(\rho)=0$ simplify to 
\[
4\sqrt{15}(\alpha-\sqrt{15}\beta)\beta + 180\beta^2\rho^2 \pm 45\sqrt{70}\beta\gamma\rho^4 - 525\gamma^2\rho^6 = 0,
\]
which is a six-degree polynomial equation with only even degrees. Then, we can make $t=\rho^2$ and get cubic equations in $t$, which we can explicitly solve.

Let us define some parameters that will help us to describe the conditions under which saddle cusps of Gauss appear.
\begin{equation}\label{eq:alphasn=6}
\begin{aligned}
&\alpha_1^\pm=\frac{-120\beta^2\pm45\sqrt{70} \beta\gamma +525\gamma^2}{4\sqrt{15}\beta},	
\quad
\alpha_2^\pm=	\frac{\beta^2}{\gamma} \sqrt{\frac27}\left(9\pm4\sqrt{3}\right),
\quad
\\
&\gamma_1^\pm=\beta\frac{\sqrt{210}\pm\sqrt{70}}{35}.
\end{aligned}
\end{equation}

%%%%%%%%%%%%%%%%%%%%%%%%%%%%%%%%%%%%%%%%%%%%%%%%%%
\begin{center}
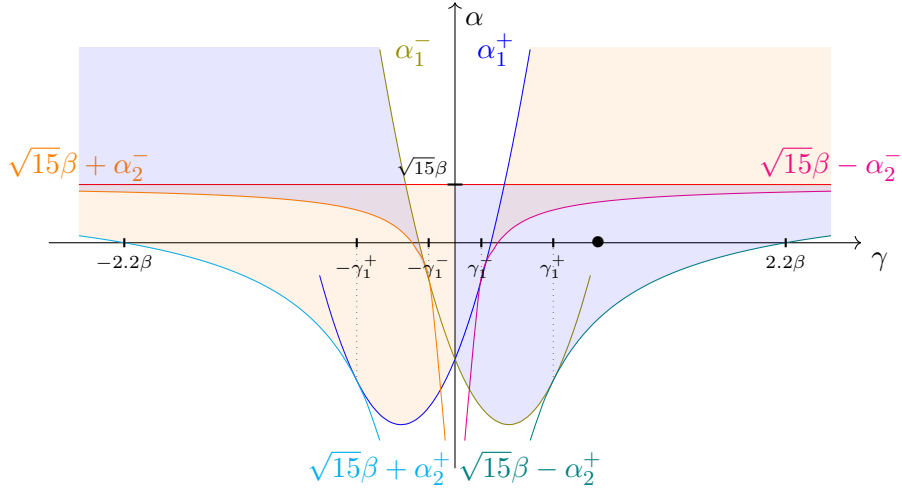
\begin{figure}[ht]
		\begin{tikzpicture}[xscale=2,yscale=0.2]
			% los ejes
			\draw[->] (-2.7,0) -- (2.7,0) node[below right] {$\gamma$};
			\draw[->] (0,-15) -- (0,16) node[below right] {$\alpha$};
			
	% Todas las funciones, sobre todo el dominio, encima para que se vean las gráficas
%
	% sqrt(15)
			\path[domain=-2.5:2.5,color=red, draw] plot[smooth] (\x,{sqrt(15)});
			
	% a2p = sqrt(15)*1/4*(-8 + 3*sqrt(70)*(\x) + 35*(\x)^2)
			\path[domain=-0.9:0.5,color=blue, draw] plot[smooth] (\x,{sqrt(15)*1/4*(-8 + 3*sqrt(70)*(\x) + 35*(\x)^2)})  node[left=2pt] {$\alpha_1^+$};
		
	% a2m = sqrt(15)*1/4*(-8 - 3*sqrt(70)*(\x) + 35*(\x)^2)
			\path[domain=0.9:-0.5,color=olive, draw] plot[smooth] (\x,{sqrt(15)*1/4*(-8 - 3*sqrt(70)*(\x) + 35*(\x)^2)})  node[right=2pt] {$\alpha_1^-$};
			
	% sqrt(15)+a1m = sqrt(15)+1.10742*1/(\x)
			\path[domain=-0.065:-2.5,color=orange, draw] plot[smooth] (\x,{sqrt(15)+1.10742*1/(\x)})  node[above] {$\sqrt{15}\beta+\alpha_2^-$};
	
	% sqrt(15)-a1m = sqrt(15)-1.10742*1/(\x)
			\path[domain=0.065:2.5,color=magenta, draw] plot[smooth] (\x,{sqrt(15)-1.10742*1/(\x)})  node[above] {$\sqrt{15}\beta-\alpha_2^-$};
			
	% sqrt(15)+a1p = sqrt(15)+(sqrt(2/7)*(9 + 4*sqrt(3)))*1/(\x)
			\path[domain=-2.5:-0.5,color=cyan, draw] plot[smooth] (\x,{sqrt(15)+(sqrt(2/7)*(9 + 4*sqrt(3)))*1/(\x)}) node[below] {$\sqrt{15}\beta+\alpha_2^+$};
	
	% sqrt(15)-a1p = sqrt(15)-(sqrt(2/7)*(9 + 4*sqrt(3)))*1/(\x)
			\path[domain=2.5:0.5,color=teal, draw] plot[smooth] (\x,{sqrt(15)-(sqrt(2/7)*(9 + 4*sqrt(3)))*1/(\x)}) node[below] {$\sqrt{15}\beta-\alpha_2^+$};

			\clip (-2.5,-14) rectangle (2.5,13);
			% funciones por trozos para hacer el relleno
			
			% sqrt(15)
			\path[domain=0:0.653085, name path=2sqrt15] plot[smooth](\x,{sqrt(15)});
			\path[domain=0.653085:2.5, name path=3sqrt15] plot[smooth](\x,{sqrt(15)});
			\path[domain=-2.5:-0.653085, name path=4sqrt15] plot[smooth](\x,{sqrt(15)});
			\path[domain=-0.653085:0, name path=5sqrt15] plot[smooth](\x,{sqrt(15)});
						
			% a2p = sqrt(15)*1/4*(-8 + 3*sqrt(70)*(\x) + 35*(\x)^2)
			\path[domain=-0.653085:0, name path = 5a2p] plot[smooth] (\x,{sqrt(15)*1/4*(-8 + 3*sqrt(70)*(\x) + 35*(\x)^2)});
			\path[domain=0.174994:2.5, name path = 6a2p] plot[smooth] (\x,{sqrt(15)*1/4*(-8 + 3*sqrt(70)*(\x) + 35*(\x)^2)});
			
			% a2m = sqrt(15)*1/4*(-8 - 3*sqrt(70)*(\x) + 35*(\x)^2)
			\path[domain=-2.5:-0.174994, name path = 1a2m] plot[smooth] (\x,{sqrt(15)*1/4*(-8 - 3*sqrt(70)*(\x) + 35*(\x)^2)});
			\path[domain=0:0.653085, name path = 2a2m] plot[smooth] (\x,{sqrt(15)*1/4*(-8 - 3*sqrt(70)*(\x) + 35*(\x)^2)});

			% sqrt(15)+a1m = sqrt(15)+1.10742*1/(\x)
			\path[domain=-2.5:-0.174994,name path=1pa1m] plot[smooth] (\x,{sqrt(15)+1.10742*1/(\x)});
						
			% sqrt(15)-a1m = sqrt(15)-1.10742*1/(\x)
			\path[domain=0.174994:2.5,name path = 6ma1m] plot[smooth] (\x,{sqrt(15)-1.10742*1/(\x)});
			
			% sqrt(15)+a1p = sqrt(15)+(sqrt(2/7)*(9 + 4*sqrt(3)))*1/(\x)
			\path[domain=-2.5:-0.653085,name path = 4pa1p] plot[smooth] (\x,{sqrt(15)+(sqrt(2/7)*(9 + 4*sqrt(3)))*1/(\x)});
			
			% sqrt(15)-a1p = sqrt(15)-(sqrt(2/7)*(9 + 4*sqrt(3)))*1/(\x)
			\path[domain=0.653085:2.5,name path = 3ma1p] plot[smooth] (\x,{sqrt(15)-(sqrt(2/7)*(9 + 4*sqrt(3)))*1/(\x)});
				
			% rellenos
			\tikzfillbetween[of=1a2m and 1pa1m, on layer=main]{blue, opacity=0.1};
			\tikzfillbetween[of=2a2m and 2sqrt15, on layer=main]{blue, opacity=0.1};
			\tikzfillbetween[of=3sqrt15 and 3ma1p, on layer=main]{blue, opacity=0.1};
			\tikzfillbetween[of=4pa1p and 4sqrt15, on layer=main]{orange, opacity=0.1};
			\tikzfillbetween[of=5a2p and 5sqrt15, on layer=main]{orange, opacity=0.1};
			\tikzfillbetween[of=6ma1m and 6a2p, on layer=main]{orange, opacity=0.1};
			
% tiks en X
			
			%	-g3p
			\draw[-,very thin, dotted ] ({-0.653085},0) -- ({-0.653085},0) node[below=1pt] {\tiny$-\gamma_1^+$};
			\draw[-,very thin, dotted ] ({-0.653085},0) -- ({-0.653085},{-9.16358});
			\draw[-,thick] ({-0.653085},0.3) -- ({-0.653085},-0.3);
	
			%	-g3m
			\draw[-,very thin, dotted ] ({-0.174994},0) -- ({-0.174994},0) node[below=1pt] {\tiny$-\gamma_1^-$};
			\draw[-,very thin, dotted ] ({-0.174994},0) -- ({-0.174994},{-2.45537});
			\draw[-,thick] ({-0.174994},0.3) -- ({-0.174994},-0.3);
			
			%	g3p
			\draw[-,very thin, dotted ] ({0.653085},0) -- ({0.653085},0) node[below=1pt] {\tiny$\gamma_1^+$};
			\draw[-,very thin, dotted ] ({0.653085},0) -- ({0.653085},{-9.16358});
			\draw[-,thick] ({0.653085},0.3) -- ({0.653085},-0.3);

			%	g3m
			\draw[-,very thin, dotted ] ({0.174994},0) -- ({0.174994},0) node[below=1pt] {\tiny$\gamma_1^-$};
			\draw[-,very thin, dotted ] ({0.174994},0) -- ({0.174994},{-2.45537});
			\draw[-,thick] ({0.174994},0.3) -- ({0.174994},-0.3);

			\draw[-,very thin, dotted ] ({2.1983},0) -- ({2.1983},0) node[below=1pt] {\tiny$2.2\beta$};
%			\draw[-,very thin, dotted ] ({0.174994},0) -- ({0.174994},{-2.45537});
			\draw[-,thick] ({2.1983},0.3) -- ({2.1983},-0.3);

			\draw[-,very thin, dotted ] ({-2.1983},0) -- ({-2.1983},0) node[below=1pt] {\tiny$-2.2\beta$};
%			\draw[-,very thin, dotted ] ({0.174994},0) -- ({0.174994},{-2.45537});
			\draw[-,thick] ({-2.1983},0.3) -- ({-2.1983},-0.3);
            
			% tiks en Y
			\draw[-,very thin, dotted ] (0,{sqrt(15)}) -- (0,{sqrt(15)}) node[above left=-1pt] {\tiny$\sqrt{15}\beta$};
			\draw[-,thick] (0.05,{sqrt(15)}) -- (-0.05,{sqrt(15)});

 % ejemplo numerico
\node at (0.95,0) (ex) {$\bullet$}; 
    \end{tikzpicture}
\caption{Regions of values for $\alpha$ and $\gamma$ (depending on $\beta$) providing saddle cusps of Gauss (case $n=6$). Blue areas correspond with saddle points with %angular coordinates 
$\theta=\pi/6,3\pi/6,5\pi/6,7\pi/6,9\pi/6,11\pi/6$. Orange areas correspond with saddle points with $\theta=0,2\pi/6,4\pi/6,\pi,8\pi/6,10\pi/6$. The black bullet corresponds with values of example \eqref{eq:6star}. Expressions for $\alpha_1^\pm, \alpha_2^\pm, \gamma_1^\pm$ are given in \eqref{eq:alphasn=6}.}
\label{fig:regn6}
\end{figure}
\end{center}

\noindent 
Doing computations with {\em Mathematica}, we get that there are saddle cusps of Gauss inside the pupil, $(\rho,\theta)$ in the following situations (see Figure~\ref{fig:regn6}):
\begin{itemize}%\itemindent35pt
\item with $\theta=\frac{\pi}{6},\frac{3\pi}{6},\frac{5\pi}{6},\frac{7\pi}{6},\frac{9\pi}{6},\frac{11\pi}{6}$ if: \smallskip

\centerline{\renewcommand{\arraystretch}{1.5}
\begin{tabular}{|>{\centering\arraybackslash}p{4cm}|>{\centering\arraybackslash}p{3.2cm}|>{\centering\arraybackslash}p{4cm}|}
\hline 
$\gamma<-\gamma_1^-$ \par $\sqrt{15}\beta + \alpha_2^-<\alpha<\alpha_1^-$
 &
 $0<\gamma<\gamma_1^+$ \par $\alpha_1^-<\alpha<\sqrt{15}\beta$
 &
 $\gamma\geqslant\gamma_1^+$ \par $\sqrt{15}\beta-\alpha_2^+<\alpha<\sqrt{15}\beta$
 \\\hline
\end{tabular}}\smallskip

\item with $\theta=0,\frac{2\pi}{6},\frac{4\pi}{6},\pi,\frac{8\pi}{6},\frac{10\pi}{6}$ if \smallskip

\centerline{\renewcommand{\arraystretch}{1.5}
\begin{tabular}{|>{\centering\arraybackslash}p{4cm}|>{\centering\arraybackslash}p{3.2cm}|>{\centering\arraybackslash}p{4cm}|}
\hline 
$\gamma>\gamma_1^+$ \par $\sqrt{15}\beta - \alpha_2^-<\alpha<\alpha_1^+$
 &
 $-\gamma_1^+<\gamma<0$ \par $\alpha_1^+<\alpha<\sqrt{15}\beta$
 &
 $\gamma\leqslant-\gamma_1^+$ \par $\sqrt{15}\beta+\alpha_2^+<\alpha<\sqrt{15}\beta$
 \\\hline
\end{tabular}}\smallskip

\end{itemize}

Here we have to observe that there are two regions where we can get twelve saddle points within two different circles, these regions are \smallskip

\centerline{\renewcommand{\arraystretch}{1.5}
\begin{tabular}{|>{\centering\arraybackslash}p{5.8cm}|>{\centering\arraybackslash}p{5.8cm}|}
\hline 
$\gamma<-\gamma_1^+$ \par $\sqrt{15}\beta+\alpha_2^-<\alpha<\min\{\sqrt{15}\beta , \alpha_1^-\}$
 &
 $\gamma>\gamma_1^-$ \par $\sqrt{15}\beta-\alpha_2^-<\alpha<\min\{\sqrt{15}\beta , \alpha_1^+\}$
 \\\hline
\end{tabular}}\smallskip
\ 
\end{proof}

As a consequence, following Proposition~\ref{propA} and Proposition~\ref{th:fertile}, we can state that the maximum $p$-fold symmetry of the star pattern is set by the number $n$.

\begin{corollary}\label{corollary}
A $p$-fold symmetric wave aberration may only generate a starburst that either has $p$ points which are uniformly distributed in a circle centered at the pupil center, or it has $2p$ points, where $p$ lay in a circle, and also the other $p$ points, but at a different distance (say shorter), with the condition that each of the long-short pairs of starburst points are aligned along the same symmetric reflection plane.
\end{corollary}
\begin{proof}
In each meridian plane, being a plane of symmetry, there are only a maximum of two starburst points. If there is only one, due to $p$-fold symmetry, there can only be $p$ starburst points. In the other case, if each plane of symmetry contains two starburst points, the same structure must be replicated in the rest of the planes of symmetry to maintain the $p$-fold symmetry, so there are $2p$ starburst points. 
\end{proof}

Finally, another important implication of the relations between $\alpha$, $\beta$, and $\gamma$, for all the studied cases illustrated in figures 1 to 4, is that the admissible values of defocus ($\alpha$ value) for the appearance of saddle cusps of Gauss are bounded. In other words, if the subject looking at a star has a certain level of defocus (for instance, because not correcting its refraction with spectacles), he or she could not see the star-like starburst.

\subsection{Relation between Zernike coefficients: necessary conditions for the appearance of starbursts}\label{sec:conditions}

Under low-light conditions (required to see starbursts), a typical value of pupil radius is $3.5$ $mm$; we use this value from now on. Some population statistical studies of Zernike aberration structure in normal and healthy eyes have revealed that higher-order aberrations tend to be randomly distributed about a mean value of zero~\cite{Thibos:02}, except for spherical aberration. Also, there is no apparent correlation between the degree of ametropia and spherical aberration~\cite{Kingston:13}. The mean value of spherical aberration ($Z_{4}^{0}$ accompanying coefficient in the Zernike expansion of the wave aberration function) is positive and around $0.18\,\mu m$ over a $3\,mm$ pupil radius ($Rp$)~\cite{Kingston:13}. Taking into account these reference values, in \eqref{eq:W} we set $\beta = 0.2\,\mu m$ using a $3.5\,mm$ pupil radius. The coefficients accompanying the Zernike terms are given in $\mu m$.
Also, as a gross approximation to spherical equivalent from the defocus Zernike coefficient ($\alpha$ in \eqref{eq:W}) is given by formula~\cite{Thibos:02}: $M = \frac{4\sqrt{3}\,\alpha}{Rp^2}$, being $R_p$ the pupil radius.

Considering the previous observations, we analyze only limited values --still wide enough to capture the totality of possible real eye cases-- of the parameters for defocus ($\alpha$ as the coefficient of $Z_{2}^{0}$), spherical ($\beta$ as the coefficient of $Z_{4}^{0}$), and $Z_{n}^{n}$ (with coefficient $\gamma$) aberration. 

Specifically, we take $\beta=0.2$, and analyze the situation for values of $\gamma$ in the appropriate interval given in the proof of Proposition~\ref{th:fertile} for the different values of $n$. In particular, in absence of defocus ($\alpha=0$), the admissible values for $\gamma$ are in the intervals:
\begin{itemize}
    \item $\gamma\in[-2.5,2.5]$ for $n=3$, 
    \item $\gamma\in[-0.76,0.76]$ for $n=4$, 
    \item $\gamma\in[-0.45,0.45]$ for $n=5$,
    \item $\gamma\in[-0.44,0.44]$ for $n=6$.      
\end{itemize}

\section{Numerical simulations}\label{sec:numerical}
In this section, we explain the algorithms used to compute the set of critical curves ($G=0$); cusps of Gauss: $\nabla G = 0$; saddle cusp of Gauss: $\nabla G = 0$ with $\det(\text{Hess}\,G )< 0$, and caustic patterns, and apply them to obtain different maps for several archetypal examples of starbursts.

\subsection{Computation of critical and singular points of the Hessian determinant and caustic patterns}

% \textcolor{red}{PROGRAMA UTILIZADO: figuresstarburstpaper.m}
We denote $(x, y)$ the coordinates of a wavefront aberration function ($W$) at the eye's exit pupil. A ray at this plane is optically propagated onto the retina, assumed as usual to be a plane.
However, in visual optics, it is better to provide angular magnitudes at the retina, denoted with $(\xi, \eta)$. These coordinates are: $\xi = x'/f$ and $\eta = y'/f$, where $(x', y')$ are the ray Cartesian coordinates at the retina plane, and $f$ is the distance between this plane and the exit pupil plane. 

The relation between $(x, y)$ and $(\xi, \eta)$ is given by~\cite{Barbero:22}:
\begin{equation}
\xi = -\frac{\partial W}{ \partial x}, \quad \eta = -\frac{\partial W}{\partial y}. 
\end{equation}\label{eq:mapping}
We note that here, contrary to what was used in reference~\cite{Barbero:22}, the defocus aberration ($Z_2^0$) is explicitly included in $W$.

From the analytical expression of $W$, we obtain algebraic expressions of $G$, and subsequently numerically evaluate it over a fine Cartesian uniform mesh within the circular domain of the eye exit's pupil. To compute the set $G=0$, we first interpolate the discrete set of data of $G$, and later obtain the iso-level curves using Matlab built-in functions \textit{contour} and \textit{getContourLineCoordinates}.

To compute the cusps of Gauss, we first numerically solve, using Matlab built-in function \textit{vpasolve}, the set of two equations provided by $\nabla G = 0$, and subsequently select those solutions that are saddle: $\det(\text{Hess}\,G )< 0$. 

To compute the caustic curves at the retina plane, we optically map the set $G=0$ with equations \eqref{eq:mapping}. Also, the same applies to obtain the location of the cups of Gauss projected to the retina plane.

\subsection{Arquetypical examples of starbursts}\label{sec:examples}
As previously mentioned, we concentrate on star-like starbursts, \emph{i.e.}, with a $p$-fold symmetry.
A $p$-fold symmetric starburst can be classified depending on the number of points. A first classification is based on the parity of the number of points, so that we can say an odd or even starburst.
A second classification concerns the number of points and their length.
There are two possibilities: $p$ equally-spaced points from the starburst center, or $2p$, where $p$ points are the same lenght, and the other $p$ points have another lenght; one set is due to cusp caustics and the other to two very close fold caustics (see hypothesis \ref{oh}). We name the first equally-distanced points starbursts and the second non-equally-distanced points starbursts. Our methodology allowed us to find four archetypical starbursts: $3$, $4$, $5$, and $6$ equally-distanced points starbursts, and an $8$ non-equally-distanced points starburst.

\begin{figure}[ht]
\centering
\includegraphics[width=12.5cm]{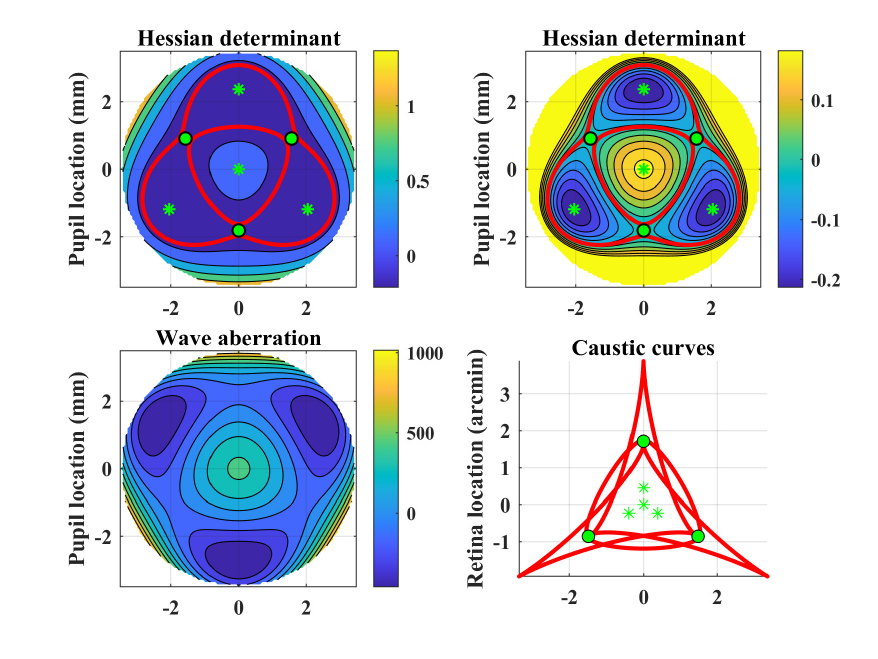}
\caption{Wave aberration $W  = 0.2 Z_{4}^{0}  + 0.2 Z_{3}^{3}$ given in \eqref{eq:3star}, Hessian determinant maps, and caustic patterns of equally-distanced three-point starburst. Green stars denote the cusps of Gauss, and green points surrounded by black circles denote the saddle cusps of Gauss in both the Hessian determinant map and its projection onto the retina plane. }
\label{fig:3stars}
\end{figure}

\subsubsection{Odd equally-distanced points starbursts}

The first example is a wave aberration given by:
\begin{equation}\label{eq:3star}
W  = 0.2 Z_{4}^{0}  + 0.2 Z_{3}^{3}.
\end{equation}

Figure \ref{fig:3stars} shows the wave aberration, Hessian determinant of $W$, and caustic pattern generated by \eqref{eq:3star}. We provide two Hessian determinant maps. In the left column, the colorbar is proportional to the Hessian determinant over all the pupil; however, in the right column, the colorbar is clipped to specific values around zero to reveal the structure of the Hessian determinant in close regions to the cusps of Gauss.

The wave aberration preserves the 3-fold symmetry (Proposition~\ref{propA}) between the Hessian determinant map and the caustic pattern. 
Overall, there are $7$ cusps of Gauss, $3$ being fertile (saddle cusps generating near fold caustics), setting the planes of specular symmetry and giving rise to a three-point starburst pattern because of the caustic cusps associated with each of the three fertile cusps of Gauss. 

\begin{figure}[ht]
\centering\includegraphics[width=12.5cm]{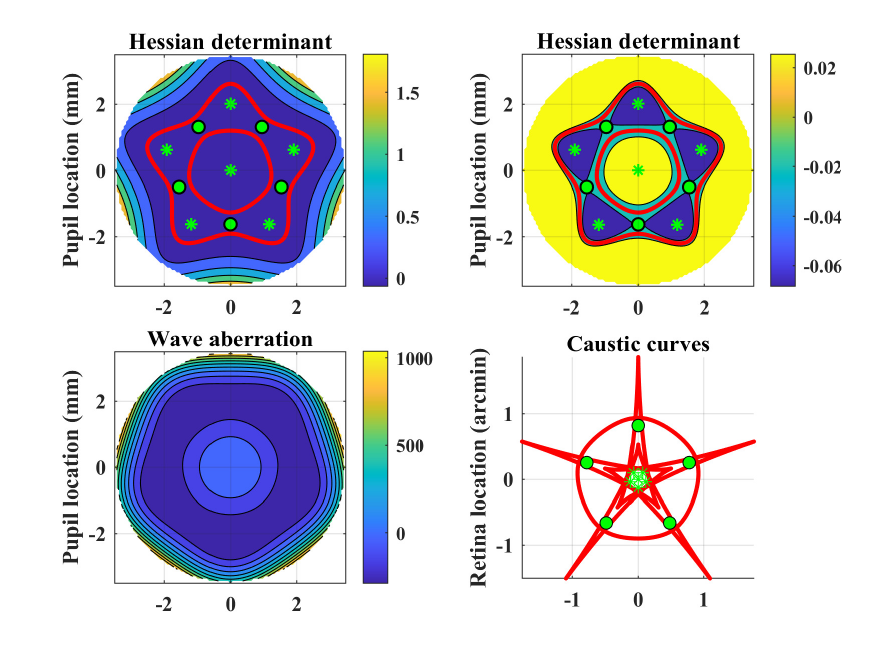}
\caption{Wave aberration $W  =  0.2 Z_{2}^{0}  + 0.2 Z_{4}^{0}  + 0.07 Z_{5}^{5}$ given in \eqref{eq:5star}, Hessian determinant map, and caustic patterns of equally-distanced five-point starburst. Green stars denote the cusps of Gauss, and green points surrounded by black circles denote the saddle cusps of Gauss in both the Hessian determinant map and its projection onto the retina plane.}
\label{fig:5stars}
\end{figure}

The second example is a wave aberration given by:
\begin{equation}\label{eq:5star}
W  = 0.2 Z_{2}^{0} +  0.2 Z_{4}^{0}  + 0.07 Z_{5}^{5},
\end{equation}
which generates a 5-fold symmetry. There are 11 cusps of Gauss, five of them being fertile points (as shown in Figure \ref{fig:5stars}). In this case, each fertile cusp of Gauss generates a cusp caustic, and the action of two opposite fertile points, two converging fold caustics; the combination of both generates a long-bright
starbust point. Overall, the caustic pattern exhibits a five-point starburst pattern.

In both previous examples, figures~\ref{fig:3stars} and~\ref{fig:5stars}, the number of starburst points equals the number of fertile cusps of Gauss, being equally-distanced odd starbursts.

\subsubsection{Even equally-distanced, and non-equal-distanced points starbursts}
A first example will be a four equally-distanced starburst obtained by a wave aberration given by:
\begin{equation}\label{eq:4star}
W  = 0.2 Z_{4}^{0}  + 0.15 Z_{4}^{4},\\
\end{equation}

\begin{figure}[ht]
\centering\includegraphics[width=12.5cm]{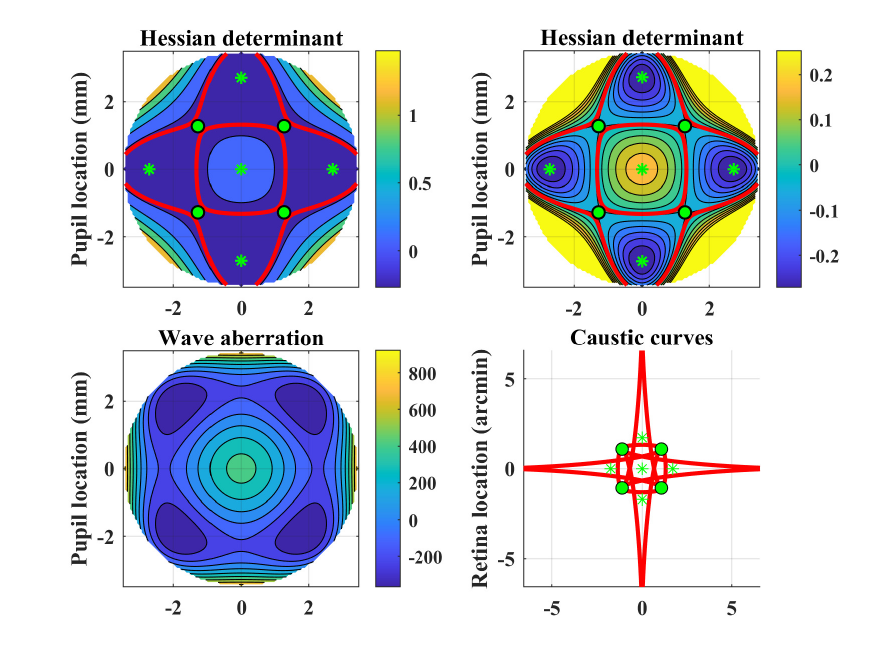}
\caption{Wave aberration $W  = 0.2 Z_{4}^{0}  + 0.15 Z_{4}^{4}$ given in \eqref{eq:4star}, Hessian determinant map, and caustic patterns of an equally-distanced four-point starburst. Green stars denote the cusps of Gauss, and green points surrounded by black circles denote the saddle cusps of Gauss in both the Hessian determinant map and its projection onto the retina plane.\label{fig:4stars}}
\end{figure}

As in the previous subsection, Figure \ref{fig:4stars} shows the wave aberration, the Hessian determinant, and the caustic pattern of equation \eqref{eq:4star}.

There are nine cusps of Gauss, $4$ of which are saddles, determining a $4$-fold caustic pattern symmetry. The cusps are arranged in pairs (alternating saddle and no-saddle pairs) along lines of reflectional symmetry. Whereas the lines including the fertile cusp of Gauss induce cusp caustics, they do not generate starburst points because they are inside the central core; the lines including the ordinary cusp of Gauss induce points obtained from the convergence of two-fold caustics (generated by opposite pairs of fertile cusps of Gauss). Overall, we get an equally distanced $4$-point starburst.

The second example is a six equally-distanced starburst provided by a wave aberration given by:

\begin{equation}\label{eq:6star}
W  = 0.2 Z_{4}^{0}  + 0.19 Z_{6}^{6},\\
\end{equation}

\begin{figure}[ht]
\centering\includegraphics[width=12.5cm]{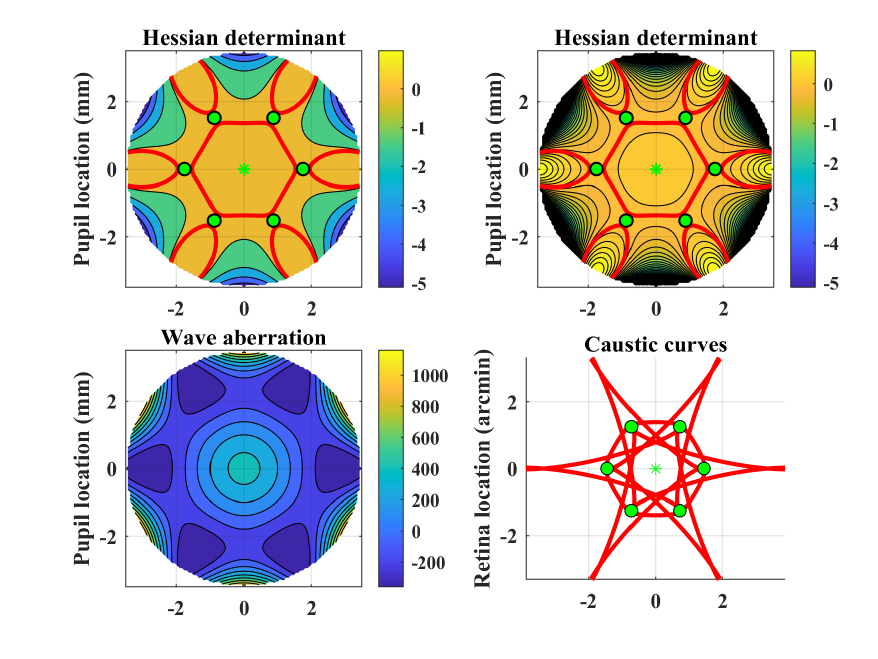}
\caption{Wave aberration $W  = 0.2 Z_{4}^{0}  + 0.19 Z_{6}^{6}$ given in \eqref{eq:6star}, Hessian determinant map, and caustic patterns of an equally-distanced six-point starburst. Green stars denote the cusps of Gauss, and green points surrounded by black circles denote the saddle cusps of Gauss in both the Hessian determinant map and its projection onto the retina plane.\label{fig:6stars}}
\end{figure}

As in the previous subsection, Figure \ref{fig:6stars} shows the wave aberration, the Hessian determinant, and the caustic pattern of \eqref{eq:6star}.

Now, the number of cusps of Gauss is $7$, $6$ of which are fertile, configuring $6$ lines of reflectional symmetry that determine a $6$-fold caustic pattern symmetry.
As in the case of the 5-point staburst, each fertile cusp of Gauss generates a cusp caustic, and the action of two opposite fertile points, two converging fold caustics; the combination of both generates a long-bright starbust point. Thus inducing an equally distant 6-point staburst.

The final example is a wave aberration generated by:
\begin{equation}\label{eq:8stars}
W  =  0.2 Z_{4}^{0}  + 0.09 Z_{4}^{4},\\
\end{equation}

\begin{figure}[ht]
\centering\includegraphics[width=12.5cm]{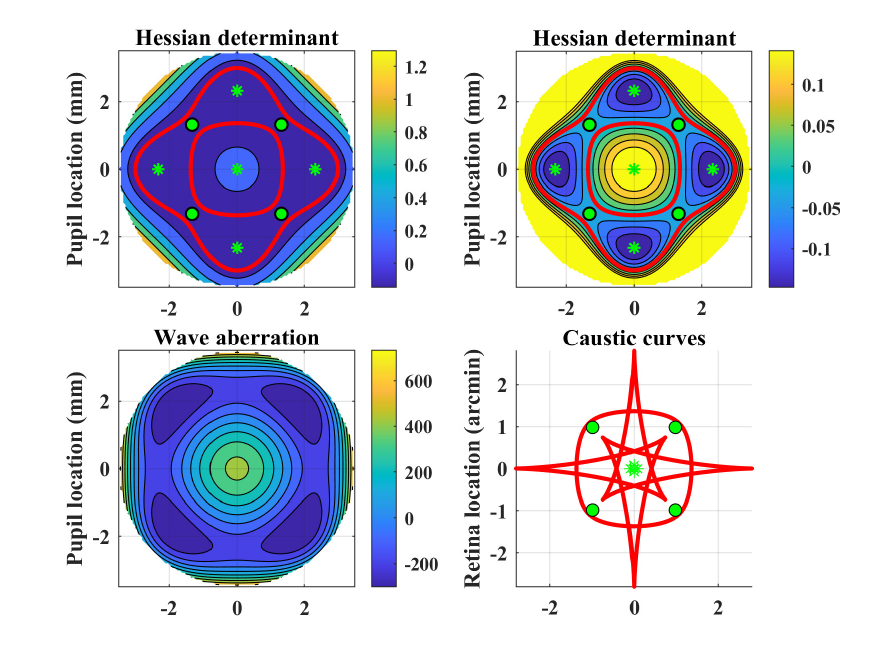}
\caption{Wave aberration $W  =  0.2 Z_{4}^{0}  + 0.09 Z_{4}^{4}$ given in \eqref{eq:8stars}, Hessian determinant map, and caustic patterns of non-equal-distance eight-point starburst. Green stars denote the cusps of Gauss, and green points surrounded by black circles denote the saddle cusps of Gauss in both the Hessian determinant maps and their projection onto the retina plane.}
\label{fig:8stars}
\end{figure}

As in the case of wave aberration \eqref{eq:4star}, the number of total and saddle cusps of Gauss is 9 and 4, respectively. Also, the cusps are arranged in pairs (alternating saddle and no-saddle pairs) along lines of reflectional symmetry. However, now the lines including the fertile cusp of Gauss induce short starburst points due to the presence of cusp caustics, and the lines including the ordinary cusp of Gauss induce long points obtained from the convergence of two-fold caustics (generated by opposite pairs of fertile cusps of Gauss). Then, the structure of caustic lines (see Figure \ref{fig:8stars}) creates not a 4-point starburst but an 8-point starburst, where 4 points are longer than the other 4. Hence, this exemplifies a non-equal-distance starburst as stated in Corollary~\ref{corollary}.

\section{Discussion}\label{sec:dis}

Knowing a subject's wave aberration, our theory enables us to predict whether somebody could see a star-type starburst pattern, and in that case, to predict an upper bound to the number of starburst points seen. 
For this to happen, a certain combination of symmetric aberrations and a $Z_n^n$ term --being them dominant over the rest of the Zernike monomials-- must exist. We note that, in some cases, there are two caustic cusps associated with each point (see figures \ref{fig:3stars}, \ref{fig:5stars}, and \ref{fig:6stars}), which presumably would induce a brighter sensation of points than in other cases.

%\subsection{Necessary and sufficient conditions}

Although we have provided the necessary conditions to perceive those types of starbursts, we warn that they may not be sufficient. Besides excessive blurring, scattering from artificial (e.g., streetlights) light may also prevent seeing starburst because there is insufficient luminance contrast~\cite{Bara:21}. This scattering may be produced either in the atmosphere or within the eye, which occurs when the retina receives light directly from artificial light sources, producing visual skyglow~\cite{Bara:23}. Still, another reason exists to perceive a \textit{star}: thin light lines radiating from the central bright spot may be formed by edge diffraction produced by fine structures within the human eye. These fine structures could, under special viewing conditions, interfere with vision, such as pupil blocking by eyelids when squinting or being present due to some permanent ocular conditions, such as cataracts, lens Y-suture, or corneal swelling. In passing, it is worth mentioning that edge diffraction induced by mirror spiders is the cause of diffraction spikes found in images recorded by reflective telescopes~\cite{Harvey, Lendermann}.

We have also provided threshold admissible defocus values, so as not to destroy the star-like starburst. On one hand, the defocus may produce a much more complex caustic pattern. For instance, a snowflake pattern (as shown in~\cite[Fig.~2]{Xu}) may occur when the saddle cusps of Gauss separate from the pupil center; on the other hand, the defocus may create too blurred patterns, even to discern a hot spot in the energy. The sensitivity of starburst patterns to defocus has also been analyzed theoretically in~\cite{Rubinstein19}. Last, but not least, if the retina starburst extension is too small, its shape may not be resolved simply because of the human eye's visual resolution. 

%\subsection{Applications wavefront reconstruction}

Reversing the argument of what we first stated at the beginning of this section, knowing how a subject sees a star ($p$-fold symmetry and the number of points), a promising application of our theoretical framework is that we could infer some basic properties of the wave aberration function of such a subject. 
Recovering some properties of the wave aberration function from the caustic pattern is an unexplored and exciting problem, somewhat mathematically connected to the so-called Hessian topology~\cite{Arnold:04}. This problem has far-reaching potential applications in the field of multifocal optical design~\cite{BarberoOE:22}, wavefront sensing~\cite{Ribak:01}, or even in electron optics~\cite{Tavabi:15}.

\section*{Acknowledgements}

S.B. thanks the support by grants PID2020-113596GB-I00 and PID2023-150166NB-I00, all funded by MCIN/AEI/10.13039/501100011033.\smallskip

A.M.D. and L.F. thank the support by grants CEX 2020-001105-M and PID2023.149117NB.I00, funded by MICIU/AEI/ 10.13039/501100011033 and ERDF A way of making Europe.

\bibliographystyle{abbrv}
\bibliography{starburstBDF2025}

\end{document}